\documentclass[fleqn,usenatbib]{mnras}

\usepackage[T1]{fontenc}

\DeclareRobustCommand{\VAN}[3]{#2}
\let\VANthebibliography\thebibliography
\def\thebibliography{\DeclareRobustCommand{\VAN}[3]{##3}\VANthebibliography}

\usepackage{amsmath}
\usepackage{amssymb}
\usepackage{graphicx}
\usepackage{float}
\usepackage{parskip}
\usepackage{caption}
\usepackage{subcaption}
\usepackage{natbib}
\usepackage{newtxtext,newtxmath}
\usepackage{lineno}

\title[What doesn't kill Gaia makes her stronger]{What doesn't kill Gaia makes her stronger}

\author[R. Arthur, A. E. Nicholson and N. J. Mayne]{
R. Arthur,$^{1}$
A. E. Nicholson$^{2}$\thanks{E-mail: arwen.e.nicholson@gmail.com}
and  N. J. Mayne$^{2}$
\\
$^{1}$Department of Computer science, Faculty of Environment Science and Economy, University of Exeter, EX4 4QL, UK.
\\
$^{2}$Department of Physics and Astronomy, Faculty of Environment Science and Economy, University of Exeter, EX4 4QL, UK.\\
}

\date{Accepted XXX. Received YYY; in original form ZZZ}

\pubyear{2024}

\begin{document}
\label{firstpage}
\pagerange{\pageref{firstpage}--\pageref{lastpage}}
\maketitle

\begin{abstract}

Life on Earth has experienced numerous upheavals over its approximately 4 billion year history. In previous work we have discussed how  interruptions to stability lead, on average, to increases in habitability over time, a tendency we called Entropic Gaia. Here we continue this exploration, working with the Tangled Nature Model of co-evolution, to understand how the evolutionary history of life is shaped by periods of acute environmental stress. We find that while these periods of stress pose a risk of complete extinction, they also create opportunities for evolutionary exploration which would otherwise be impossible, leading to more populous and stable states among the survivors than in alternative histories without a stress period. We also study how the duration, repetition and number of refugia into which life escapes during the perturbation affects the final outcome. The model results are discussed in relation to both Earth history and the search for alien life.
\\
\end{abstract}


\section{Introduction}\label{sec:intro}

The long history of life on Earth is marked by a number of `revolutions' \citep{lenton2013revolutions}, large changes in ecosystems and biogeochemical reaction networks. For example, the evolution of oxygenic photosynthesis fundamentally changed the surface chemistry of our planet, leading to the dramatic build up of oxygen in the atmosphere roughly 2.5 billion years ago, known as the Great Oxidation Event. This build up of oxygen not only caused widespread poisoning to life that had not yet adapted to an oxygen rich environment, but is also thought to have triggered a global glaciation event \citep{sahoo2012ocean,lenton2013revolutions}. Much later, the expansion of plants on land, significantly increased chemical weathering and therefore reduced atmospheric $CO_2$ \citep{lenton2012first,porada2016high}. This is hypothesised to have triggered a global glaciation and subsequent (Late Ordovician) mass extinction. Non-biotic perturbations such as changes in volcanism, asteroid impacts, plate tectonics and climate change have also caused widespread extinctions and resulted in the emergence of new ecosystems \citep{bond2017causes}. Even in cases where the ultimate cause is non-biotic e.g. asteroid impacts \citep{alvarez1980extraterrestrial} or volcanism \citep{campbell1992synchronism}, it is still often the case that life participates in positive feedback loops which worsen conditions and accelerate the extinction event \citep{bond2017causes, dal2022environmental}.

These periods of stress and their associated mass extinctions are thought to pose a problem for the body of work known as Gaia theory, which posits that life interacts with the non-living Earth so as to maintain and even improve conditions for life \citep{lovelock1974atmospheric}. Arguments against Gaia often contrast the homeostatic mechanisms currently observed against runaway feedback enhanced by life before or during extinction events, as well as particular instances where some species (or group of species) acts to worsen conditions for itself \citep{kirchner2002gaia,kirchner2003gaia}. Taken to the extreme, some have argued that life is in fact self-destructive \citep{ward2009medea}. The question of whether life is self-reinforcing or self-destructive clearly has profound implications not only for our understanding of Earth history but also for the search for life on other worlds. The answers will shape our expectations for the prevalence of life, especially complex life, in the universe. In particualr, for exoplanets (planets outside the solar system), a deeper understanding of the life-climate interaction and the impact of large-scale perturbations may well prove vital in selecting the few targets for which expensive and intensive follow-up biosignature observations are performed. 

In previous work \citep{arthur2017entropic,lenton2018selection,arthur2022selection,arthur2023dice} we have taken the optimistic position that these revolutions are a feature, not a bug. Over the course of Earth history there have been numerous cycles of extinction and recovery. We propose that these cycles should be considered as part of Gaia. In particular they are the mechanism by which Gaia can make large jumps in complexity. We refer to this mechanism as Sequential Selection with Memory or the Entropic Ratchet, which is summarised by the following three points:
\begin{enumerate}
    \item Gaian homeostasis can be destabilised  by the evolution of new species.
    \item These events cause some, or all, of the `core' or `keystone' species of the global ecosystem to go extinct. New core species then arise, which create new niches and participate in  biogeochemical cycles \citep{alroy2008dynamics}.
    \item These extinctions are not total, for example, core species can survive but become too rare to fulfill their prior ecosystem function \citep{hull2015rarity}. Successive resets therefore begin from a higher base diversity. This means the post-event biodiversification, that fills the ecological space opened by the mass extinction, builds and expands on evolutionary innovations of the preceding period, which tends to result a higher complexity and diversity of life.
\end{enumerate} 
Points one and two are likely uncontroversial, point three is why we claim this process is Gaian. After each event, global biomass and diversity could decrease or increase relative to the previous baseline, any particular event must be analysed and understood individually. However, as we have argued elsewhere \citep{arthur2022selection, arthur2023dice} there is a general tendency for systems with `memory' to increase in complexity over time, a point often noted in complex systems theory \citep{anderson2004evolution}. For Gaia, this memory is the global biota, which is reduced but not eliminated by these resets. We claim that repeated resets lead to a trend of increasing diversity and abundance together with a reduction in the rate of mass extinction, which makes the Earth (or any inhabited planet) `more Gaian' over time.

In previous work we have demonstrated this mechanism operating in a general model of co-evolution \citep{arthur2017entropic,arthur2022selection}. This mechanism also has support 
from the literature on mass extinction and biodiversity over geological time. For example \citep{benton1995diversification, newman1999extinction} demonstrate a trend over the 
Phanerozoic (the last $\sim 540$ Mya) of increasing biodiversity, despite numerous mass extinctions. For particular mass extinction events there is some evidence of increases in 
post extinction complexity. For example, from the relative abundance distribution of marine fossils \cite{wagner2006abundance} finds that complex ecosystems are more common during 
the Meso-Cenozoic (later Phanerozoic) than during the Paleozoic (earlier Phanerozoic) where the boundary is marked by Permian-Triassic mass extinction. Similar ideas have been 
discussed in the literature on mass extinction, notably the idea of `Earth System Succession’ \citep{hull2015life}. Our claim is that these extinction events, which greatly disrupt 
contemporary biota are, on average, positive for life in the long run, that is, over spans of time measured in 10s or 100s of millions of years which may incorporate a number of such 
events.

Since life has a profound effect on (bio)geochemical cycles, large scale disruptions of global ecosystems can impact these cycles. The way in which life recovers after such events can be complex \citep{sole2002recovery, sole2010simple} but these disruptions open the possibility for large changes in biogeochemical regulation that would not be possible without such events. To select some major examples
\begin{itemize}
    \item The Great Oxidation Event caused tremendous damage to existing anaerobic species, which, for the most part, could not tolerate a high oxygen environment, and also precipitated a possible Snowball Earth period (Huronian glaciation). However, the evolution of Eukaryotic and multicellular life was enabled by the higher oxygen concentrations which allowed for aerobic respiration to become dominant resulting in far more available energy for life \citep{mills2022eukaryogenesis}.
    \item The Cryogenian/Snowball Earth period at the end of the Proterozoic, precipitated by the Neoproterozoic Oxygenation Event, would have certainly had a negative effect on extant life.  Afterwards we see recovery and (enabled by higher oxygen levels) the further complexification and diversification of life, culminating in the Ediacaran biota \citep{narbonne2012ediacaran}.
    \item The causes of the End-Ediacaran extinction are uncertain, with explanations ranging from a standard mass extinction event to a more gradual biotic replacement \citep{laflamme2013end,darroch2018ediacaran}. Whatever the causes, the mass disappearance of the Ediacaran biota was immediately followed by the Cambrian explosion.
    \item During the Phanerozoic there have been numerous mass extinction events, with scholarship mostly focusing on recovery over shorter timescales. The End-Ordovician extinction is one example. Preceded by the `Great Ordovician Biodiversification Event' (GOBE) and the colonisation of land by plants \citep{lenton2012first}, the result was the increase of atmospheric oxygen and fire mediated feedbacks to stabilise atmospheric oxygen concentrations at $\sim 20\%$ \citep{lenton2016earliest}. This represents an increase in the complexity of the biogeochemical feedback network. Land plants spreading also accelerated the rate of silicate-weathering and thus increased the rate of carbon dioxide removal from the atmosphere leading to an overall cooler climate \citep{Berner1998}.
\end{itemize}
We do not claim that all mass extinction events in Earth history need have a positive impact on species diversity or abundance. For example, despite the findings of \cite{wagner2006abundance} diversity levels reached during the GOBE took tens of millions of years to recover after the largest mass extinction event of the Phanerozoic, the End-Permian \citep{raup1982mass, rohde2005cycles}. Other events are just `blips' on Gaian timescales. For example, the Cretaceous–Paleogene event, generally agreed to be caused by an asteroid impact \citep{alvarez1980extraterrestrial}, appears to have had little long term impact on trends in biodiversity \citep{rohde2005cycles}, extinction rate \citep{benton1995diversification} or global temperature \citep{scotese2021phanerozoic}. 

The complexity of the global ecosystem and life-environment feedback (i.e. Gaia) has increased over geological eons. While the consequences of complexity in ecology are debated \citep{landi2018complexity}, our view, supported by our models of co-evolving ecosystems \citep{arthur2017entropic, arthur2022selection, arthur2023dice}, is that an increase in complexity is associated with an increase in habitability and stability. In real ecologies complexity, realised through biodiversity, can enhance stability in a number of ways, from functional redundancy \citep{rosenfeld2002functional} buffering against local extinctions to the stabilization of global biogeochemical cycles. An example of the latter is the evolution of land plants which increased the efficiency of silicate weathering and, due to their evolutionary adaptation to different climate, temperature, topography etc. these factors therefore exert less influence on the silicate weathering cycle \citep{lenton2016earliest,payne2020evolution}. 

There is less work on how the total abundance of life has changed over time. Marine environments do show increased abundance over geological time \citep{bambach1993seafood, martin1996secular, allmon2014seafood} and the evolution of plants resulted in enormous increases in the mass of Earth's biota \citep{mcmahon2018deep}. Complex systems of recycling can also increase abundance, allowing limiting nutrients to `go further' than would otherwise be possible, e.g. Phosphorus is the limiting nutrient in most aquatic ecosystems where recycling ratios of around 46:1 are quoted in \cite{wilkinson2023fundamental}.

Diversity, abundance and stability should be part of any definition of `planetary habitability'. According to our framework, increases in these features are ultimately caused by life interacting with life and the planet in a way that is ultimately conducive to life, even if disasterous in the short term. Hence our identification of this mechanism as a Gaian process. Gaia, like life itself, should not be expected to have emerged \emph{de novo}, fully formed and functional, nor to be eternally static. It is reasonable that Gaia can and should evolve. 

Of more relevance to astrobiology than Earth history (where we know that life survived all previous extinction events) is the idea of Selection by Survival (SBS) \citep{ford2014natural, bouchard2014ecosystem, lenton2018selection, arthur2023gaian, arthur2023dice}. This is the obvious fact that only those planets where life survives mass extinctions have life! This could mean that the surviving life on those planets has some special properties, or the events on the surviving planets were less severe. In an astrobiological context this simpler mechanism operates alongside the Entropic Ratchet idea discussed above. Regular mass extinctions drive long term increases in habitability, but are also opportunities to lose the game entirely and end up with a planet devoid of life. For this reason we study the interplay between these two processes in our model ecosystems, and try to understand what this means for searches for extra-terrestrial life.

The detection and subsequent characterisation of exoplanets, planets orbiting stars other than the Sun, has provided a vast number of potential candidates for non-Earth biospheres. In fact, a major goal of the field of exoplanet research is to potentially detect a `biosignature' \citep{catling2018} in the atmosphere of an exoplanet \citep[e.g. the LIFE mission][]{quanz2022}. For this case, a biosignature must be detectable, requiring life to both have survived previous extinction events and established a large-scale interaction with the climate, i.e. have created an `exo-Gaia' \citep{nicholson2018exogaia}. Although the number of potential exoplanets is vast the resources required to perform detailed observations, modelling and analysis of a potential biosignature mean that targets will have to be carefully selected. In this instance, Selection by Survival (SBS) is clearly a key mechanism, where we are primarily interested in only the planets where life persists. In an astrobiological context this simpler mechanism operates alongside the Entropic Ratchet idea discussed above. Regular mass extinctions drive long term increases in habitability (the ratchet) and therefore detectability. They are also opportunities to end up with a planet devoid of life (SBS). For this reason we study the interplay between these two processes in our model ecosystems and try to understand what this means for searches for extra-terrestrial life.

In this paper we study the impacts of large scale exogenous perturbations in a model of a planetary ecosystem over geological time. Most of the great extinction events in Earth history are thought to have arisen from a combination of abiotic and biotic factors. In previous work \citep{arthur2017entropic, arthur2022selection} we only considered biotic effects, the disruption of a stable period by the evolution of new species. Here we also introduce abiotic effects, captured as a disruption of a stable period by a sudden decrease in carrying capacity. In Section \ref{sec:refugia} we review the idea of refugia (a location supporting an isolated population during some period of environmental stress),
and introduce our model in Section \ref{section:TNM_intro}. We study, in detail, the effect of a single perturbation in Section \ref{sec:one}. We extend this in Section \ref{sec:other} to look at the effect of the duration of the perturbation, the effect of repeated perturbations, and compare outcomes where there is one big refugium to a number of smaller ones. We conclude in Section \ref{sec:discussion} with a discussion of the implications of these results for Astrobiology.

\section{Refugia}\label{sec:refugia}

As discussed previously in this work and others \citep{arthur2022selection} gradually increasing diversity is the key to increasing habitability and that this can be maintained by various means which we referred to as Gaia's memory. One form this memory takes is as \emph{refugia}, areas of tolerable conditions amidst an uninhabitable environment. Refugia originally referred to the restricted ranges of various species during glacial maxima, particularly during the last ice age  \citep{stewart2001cryptic}. This body of work studies and identifies macro-refugia \citep{ashcroft2010identifying} e.g. lower latitudes that avoid glaciation and cryptic/micro-refugia \citep{stewart2010refugia} e.g. temperate areas within glacial zones. This is achieved through analysis of the pollen record \citep{bennett1991quaternary} or through genetic evidence \citep{cheddadi2006imprints}, since periods in a refugium usually correspond to a population bottleneck that leaves a distinct signature in the modern species’ DNA. 

The most extreme, planetary glaciations, snowball Earths, are also posited to have had refugia. Although most research supports the existence of open ocean conditions at equatorial and lower latitudes \citep{hyde2000neoproterozoic, peltier2004climate, song2023mid} which would act as refugia, some models \citep{braun2022ice} predict a hard snowball. In such cases microrefugia are still a possibility. Notable examples of such microrefugia are suggested by \citep{campbell2011refugium, campbell2014refugium} who claim that narrow seas (like the modern Red sea) could provide refugia for photosynthetic eukaryotic algae. Windblown dust can lower the albedo of glaciers \citep{abbot2010mudball} which can lead to pockets of liquid water called cryoconite holes. The same process acting on a larger scale on snowball earth could have created refugia for eukaryotes tolerant of cold water, low salinity, and strong radiation \citep{hoffman2016cryoconite}. Similarly, areas of “dirty ice” \citep{hawes2018dirty} could have provided stable and nutrient rich micro-refugia, as they do presently in Antarctica. \cite{lechte2019subglacial} propose that the mixing zones of oxygen rich glacial meltwater with iron rich seawater could have provided sufficient energy for chemosynthesis and therefore represent another type of refugium. Hot springs have also been proposed as providing `Noah's Arks' for photosynthetic life during hard snowball Earth events \citep{Schrag2001, Costas2008}.

A number of different types of refugia are described by \cite{bennett2008we}, demonstrating a diversity of ways that species can survive periods of climatic stress by altering their abundance and distribution. Of most relevance for us are the classical and tropical refugia, when species restrict their range to one (classical) or many (tropical) small areas in an otherwise inhospitable environment. We also note that a refugium is usually species specific \citep{stewart2010refugia} and so a refugium for one species may be inhospitable for another. This work does not seek to model any particular glaciation or other event from Earth history. Rather we aim to study how in general retreat to refugia affects the long term habitability and hence chance of life detection on any planet. Therefore we adopt a somewhat broader concept of refugia \citep{keppel2012refugia} applying the concept to whole ecosystems, having in mind something like the open equatorial ocean or habitable narrow seas during snowball Earth events.

We also note that while there has been much work on the recovery of diversity after mass extinctions e.g. \citep{condamine2013macroevolutionary}, there has been less emphasis on the recovery of biomass or productivity, reflecting a general emphasis in ecology which tends to be more interested in diversity than abundance \citep{wilkinson2023fundamental}. However the role of abundance is crucial for understanding ecosystem function \citep{spaak2017shifts}. Indeed, \cite{hull2015rarity} makes the interesting point that it is not strictly necessary for a species to go extinct, rather `keystone’ species or groups of organisms which participate in key biogeochemical cycles can fall below the abundance threshold required for them to effectively perform their roles in the cycle, see also \cite{avolio2019demystifying}. Biosphere abundance is also particularly important when searching for signs of life on distant planets; in order for life to be remotely detectable it must exist in sufficient quantifies to influence its planet in a significant way \citep{Seager_2013}.

In summary, refugia represent a way for life to survive during inhospitable conditions. At least some refugia are necessary so that life as a whole doesn't die out. These refugia act as a memory and storehouse of genetic diversity for Gaia. We will therefore be interested in understanding how the number and type of refugia interact with the Entropic Ratchet and SBS effects and what this means for the probability of complex life on a planet.

\section{The Tangled Nature Model}\label{section:TNM_intro}

The Tangled Nature Model (TNM) \citep{christensen2002tangled, jensen2018tangled} is a framework for understanding co-evolving species. The TNM is characterised by periods of stability where  groups of species persist for a significant time, often called a \emph{quasi-Evolutionary Stable Strategy} or qESS, interrupted by `quakes', where the `core' of the species network is disturbed by a newly evolved species and is rearranged or collapses completely. After a quake the system find a new qESS and the total population and composition of the biosphere drastically changes. These quakes are an inherent feature of the TNM and require no external perturbation.

One of the key characteristics of the TNM is the tendency for the biosphere to increase in total population, diversity, and stability over time. At later times the TNM biosphere is more robust and less prone to quakes and thus the periods of stability get longer as time goes on. The quakes in the model are not a deterrent to this increasing stability but rather the mechanism by which this is achieved. Previous work \citep{piovani2016linear, arthur2017entropic} has demonstrated how the TNM model is closely related to the logistic model of population dynamics. Thus, because the TNM arises from consideration of very general principles it is reasonable to posit that the model results are also of wide applicability.

In the original formulation of the TNM a parameter $\mu$ represents the `abiotic' carrying capacity and remains constant throughout experiments, while the growth rate of species within the biosphere depends on $\mu$ as well as the other species extant at that time. Later work \citep{arthur2017tangled} allowed species to directly impact the carrying capacity of the system and demonstrated that species-environment co-evolution leads to TNM biospheres tending to increase the abiotic carrying capacity over time. 

The TNM has been described numerous times in great detail \citep{christensen2002tangled, arthur2017tangled, arthur2022selection} and the reader is referred to these works for a more detailed specification of the model. Briefly, species $i$ are labelled by a length $L$ binary genome. The population of species $i$ is $N_i$ and the total population of all species is $N = \sum_i N_i$. Each species has a fitness ,$f_i$, which depends on the other extant species in the model given by
\begin{equation}\label{eqn:fitness}
    f_i = C \sum_j J_{ij} \frac{N_j}{N} + \sigma \sum_j K_{ij} N_j - \mu_0 N - \nu N^2
\end{equation}
$C, \sigma, \mu_0, \nu$ are constants and the sums are over all extant species. $J_{ij}$ is a matrix of direct inter-species interactions, $K_{ij}$ is a matrix of species environment interactions (the effect of $j$ on the environment of $i$). The values of $J$ and $K$ are chosen at random from a standard normal product distribution used for reasons of computational convenience \citep{arthur2017tangled} where a fraction of the entries, $1-\theta_J$ and $1-\theta_K$ respectively, are set to 0. $\mu_0$ is the reciprocal of the total carrying capacity, while $\nu$ is a very small damping factor which is irrelevant except in very rare cases of extremely high populations. Setting $\sigma = 0, \nu = 0$ corresponds to the original TNM of \cite{christensen2002tangled}, nonzero values give the version where species affect the environment proposed in \cite{arthur2022selection}.

\begin{table}
    \centering
    \begin{tabular}{|c||c|}
    \hline
       $C$  &  100\\ \hline
       $\theta_J$  & 0.25\\ \hline
       $\sigma$ & 0.01 \\ \hline
       $\theta_K$ & 0 \\ \hline
       $\mu_0$ & 0.1 \\ \hline
       $\nu$ & $5 \times 10^{-6}$ \\ \hline
       $p_k$ & 0.2 \\ \hline
       $p_{mut}$ & $0.01$ \\ \hline
       $L$ & 20 \\ \hline
    \end{tabular}
    \caption{TNM parameter values.}
    \label{tab:tnmparams}
\end{table}

The model consists of repeating the following steps
\begin{itemize}
\item Select an individual at random and kill it with probability $p_k$
\item Select an individual and reproduce it with probability $p(f_i) = \frac{1}{1 + e^{-f_i}}$
\end{itemize}
Each reproduction copies the binary genome of the individual with probability $p_{mut}$ to flip one of the digits, potentially creating a new species. The timescale for the model is measured in `generations' which consist of $N / p_k$ repetitions of the two steps above. We use standard values for the various parameters, listed in Table \ref{tab:tnmparams}. 

In summary, the TNM, as employed in this work, tracks populations of species which interact with each other and their environment with each timestep providing random death, reproduction and mutation of individuals. In general, TNM systems evolve to higher complexity and higher stability, moving through increasing stable periods (qESS) disrupted when the core of the life network is disrupted through the natural evolution of destabilizing species (quakes). In this work, we further add an `external' perturbation where we reduce the overall carrying capacity of the environment, representing, for example, some change in the environmental context. We perform experiments with one perturbation or multiple, where life survives in one or multiple \emph{Refugia}. Perturbations are implemented by abruptly changing the value of $\mu_0 \rightarrow \mu$ for a set number of generations and then reverting to $\mu_0$ when the stress period is over. All other parameters remain fixed during the perturbation. Higher values of $\mu$ correspond to worse conditions for life and vice versa.

\section{One Refugium}\label{sec:one}

\begin{figure*}
    \centering
    \includegraphics[width=\textwidth]{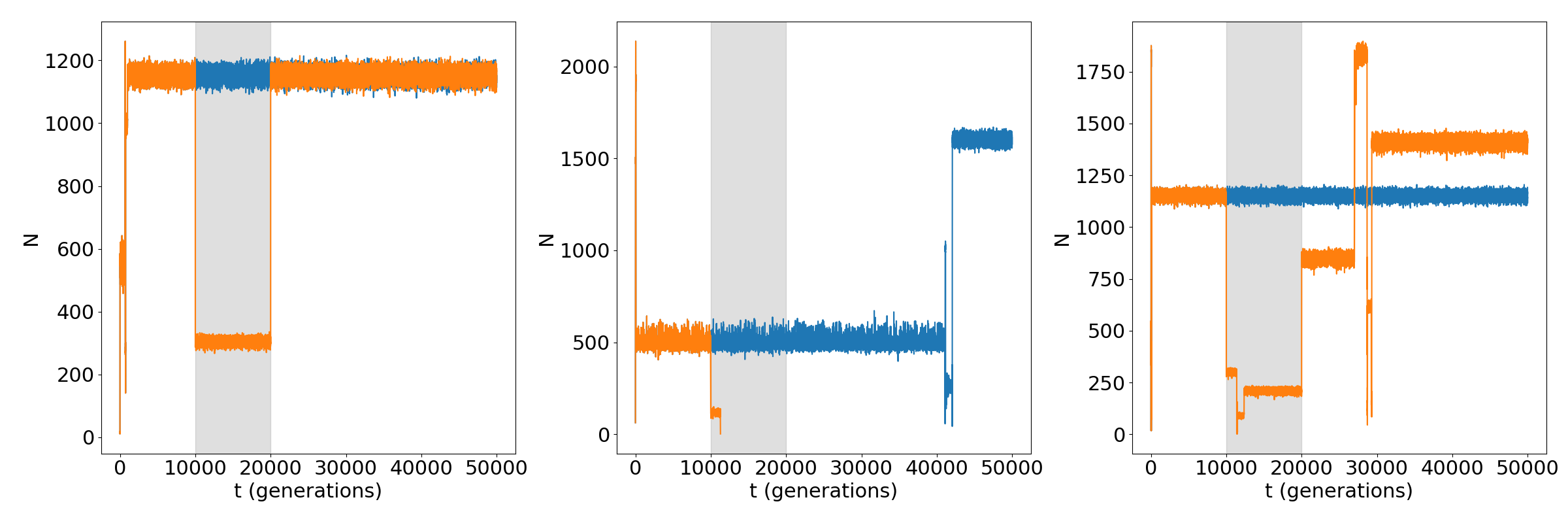}
    \caption{Some examples of TNM runs where we compare cases without perturbation (blue) to cases with perturbation $\mu_0 \rightarrow \mu = 0.4$ (orange). The shaded grey area is where the orange runs experience the perturbation. From left to right we see, minimal effect, perturbation induced extinction and perturbation induced divergence.}
    \label{fig:singleruns}
\end{figure*}

Our first experiment is similar to one preformed, in a very different context, by \cite{arthur2017tangled}. We allow the model to run as normal for $10^4$ generations then abruptly increase the value of $\mu$, run the model at that value for $10^4$ generations, then reset it to the original value and continue for another $3 \times 10^4$ generations. Some illustrative examples of different model runs are shown in in Figure \ref{fig:singleruns}. We can conceptualise this period of lower $\mu$ as something like Snowball Earth, where the planet can support a much lower abundance of life which persists in a refugium . The examples in Figure \ref{fig:singleruns} have been chosen deliberately to illustrate the most important possible results of a perturbation: no long term effect, perturbation induced extinction and perturbation induced divergence.

\begin{figure*}
    \centering
    \includegraphics[width=\textwidth]{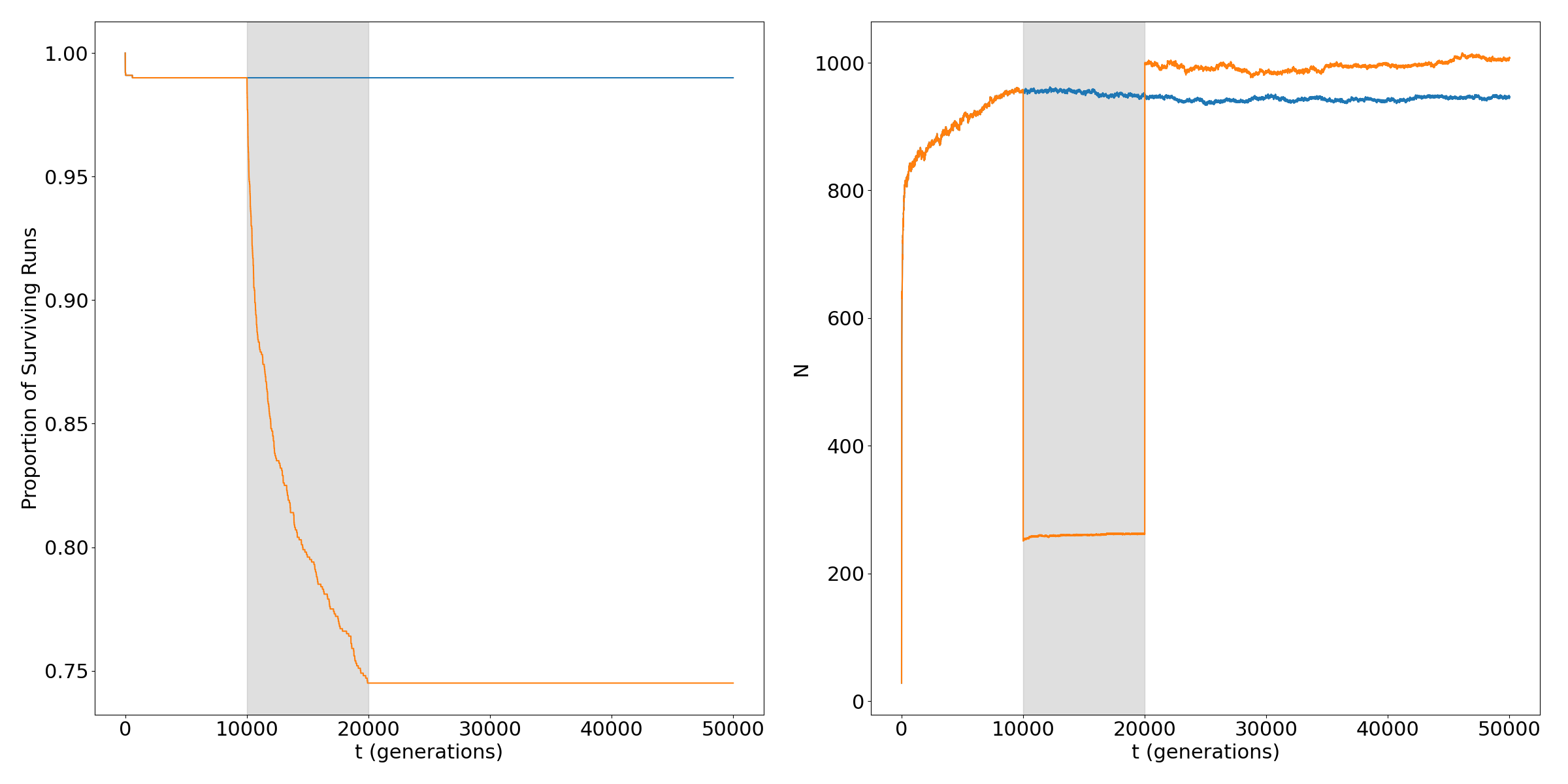}
    \caption{Averages without perturbation (blue) and with perturbation $\mu_0 \rightarrow \mu = 0.4$ (orange). The shaded grey area is where the orange runs experience the perturbation. Left shows the proportion of the $N=1000$ runs which have survived up to that point. Right shows the average population \emph{of the runs which survive the whole experiment}.}
    \label{fig:average04}
\end{figure*}

Figure \ref{fig:average04} summarises 1000 realisations of the model, for one particular value of the perturbation $\mu = 0.4$. The number of runs experiencing complete extinction of all individuals increases due to the perturbation. Figure \ref{fig:average04} also shows the average population of the subset of runs which survive to the end of the perturbed and unperturbed experiments. Notably, after the perturbation there is a jump in the average population above the unperturbed baseline which persists until the end of the run. The fact that bad conditions make total extinction more likely is quite intuitive. What is less intuitive is that runs which survive are `better off' when they experience a perturbation than they otherwise would have been. 

\begin{figure*}
    \centering
    \includegraphics[scale=0.89]{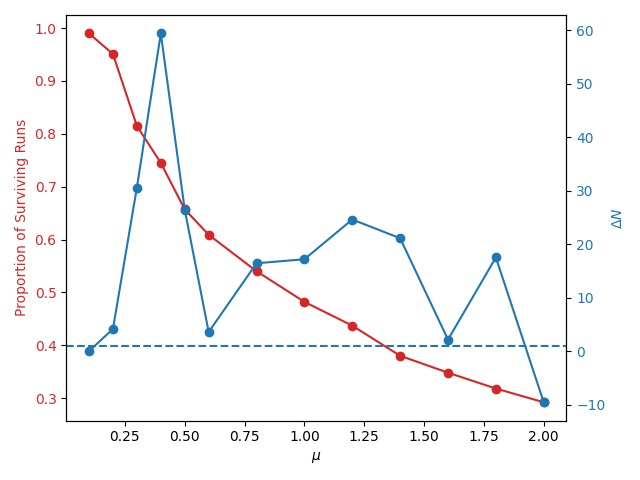}
    \caption{Proportion of runs which survive the whole experiment and ratio of population at the end of the surviving runs, with and without perturbation. Population of each run is measured by averaging the last 500 generations.}
    \label{fig:versusmu}
\end{figure*}

Figure \ref{fig:versusmu} shows the number of surviving runs as well as the difference in population between perturbed and unperturbed models at $t = 5\times10^4$ as a function of $\mu$, the perturbation size. Expectedly, as the perturbation gets stronger, more runs experience total extinction. Less expectedly, the surviving runs where there has been a perturbation have higher final populations than ones which don't. There also seems to be a peak in the response to perturbation, with a maximum around $\mu = 0.4$. Note that since the average population increases approximately logarithmically  \citep{becker2014evolution} i.e. very slowly, a 5\% increase in population is quite significant and represents a leap forward by many thousands of generations.

To understand what is happening first we note \citep{becker2014evolution, arthur2022selection} that for a mutant species, $a$, to disrupt a qESS requires it to have high enough fitness to have significant reproduction probability, i.e. the species fitness should be above a minimum value ($f_{min}$) set by,
\[
f_a > \log(\frac{p_k}{1-p_k}) = f_{min}.
\]
Using equation \ref{eqn:fitness} this means
\begin{equation}\label{eqn:barrier}
C \sum_j J_{aj} \frac{N_j}{N} + \sigma \sum_j K_{aj} N_j > f_{min} + \mu N + \nu N^2.
\end{equation}
We have set $\sigma$ and $\nu$ to be quite small and the main requirement is that the new species growth rate, $r_a = C \sum_j J_{aj} \frac{N_j}{N}$ is large enough to overcome the `barrier' on the right hand side of equation \ref{eqn:barrier} which is primarily set by the value of $\mu N$. In a qESS
\[
\frac{dN}{dt} \simeq 0.
\]
Using the mean field approximation from \cite{arthur2022selection}, and neglecting $\sigma$ and $\nu$, gives
\[
N \simeq \frac{r}{\mu},
\]
for the population in equilibrium where
\[
r = C \sum_{ij} \frac{N_i}{N} J_{ij} \frac{N_j}{N}.
\]
A sudden increase in $\mu$ will not directly affect the species composition of the TNM system, so the value of $r$ will be roughly the same immediately after the perturbation. Simply, the increase in $\mu$ will be compensated for by a decrease in $N$, while keeping the relative proportions of each species almost unchanged and thus leaving the barrier height $r$ unchanged. Given the reduced rate of reproduction, one might therefore expect \emph{fewer} quakes and since, as argued above, quakes are what drives the TNM to better (higher $N$) equilibria, the results of Figure \ref{fig:average04} and \ref{fig:versusmu} are at first glance puzzling.

\begin{figure*}
    \centering
    \includegraphics[width=\textwidth]{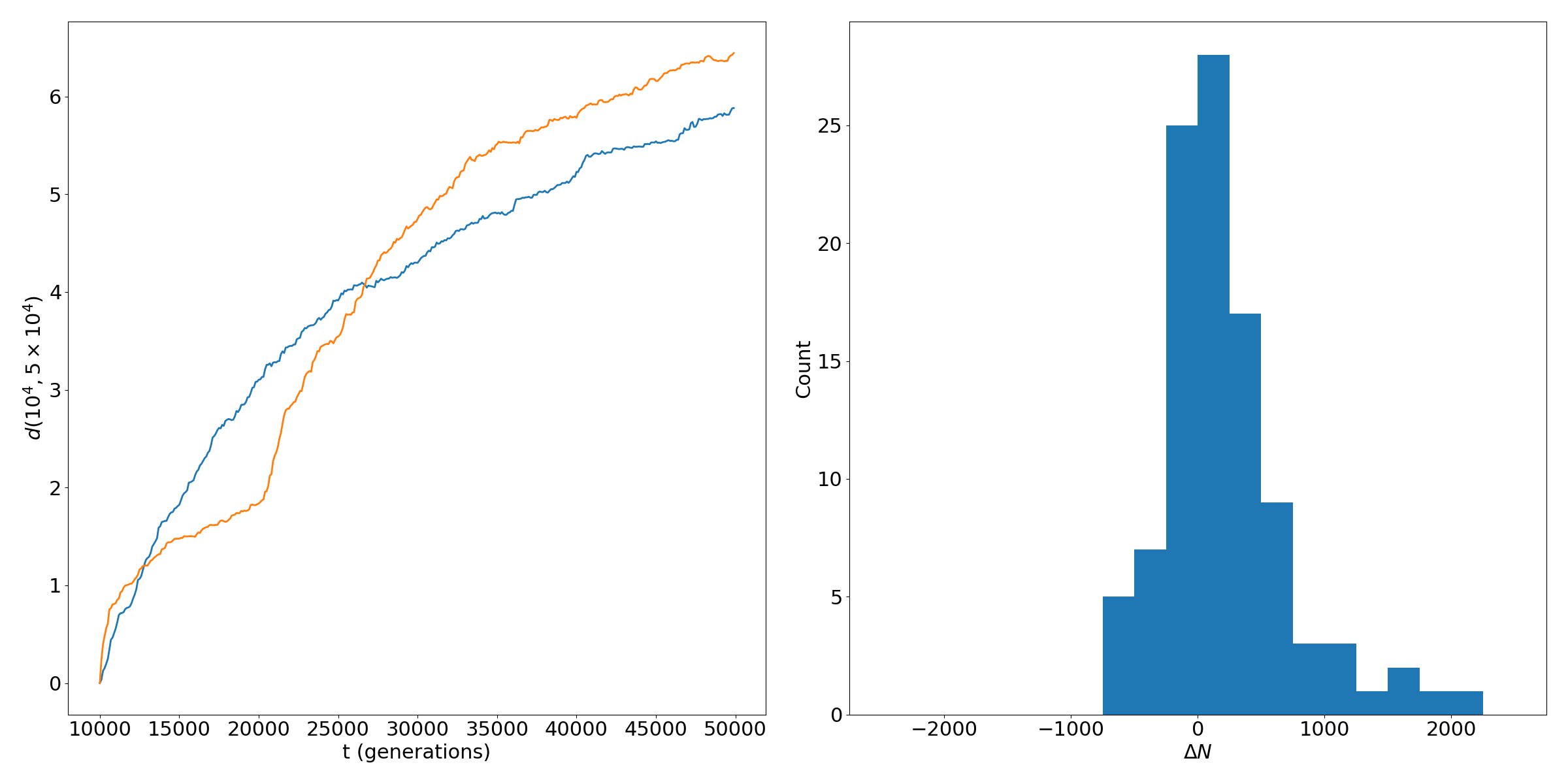}
    \caption{Showing the $\mu=0.4$ perturbation. Left shows the `genetic distance' between the core just before the perturbation and at the end of the run with (orange) and without (blue) a perturbation. Right shows the population excess among survivors of the perturbed case \emph{when there has been a quake during the perturbation}. }
    \label{fig:quakehist}
\end{figure*}

Figure \ref{fig:quakehist} shows the genetic `distance' between cores at $t_1=10^4$ (just before the perturbation) and $t_f=5\times10^4$, measured by
\begin{equation}
  d(t_1, t_f) = \sum_{i} \min_j H(C(t_1)_i, C(t_f)_j),
\end{equation}
where $C(t)$ refers to the set of core species' genomes at generation $t$ and $H$ is Hamming distance. $d$ measures the smallest number of mutations required to get from the core at $t_1$ to the core at $t_f$. The perturbed curve (orange) shows some interesting features. Just after the perturbation, the distance increases - corresponding to core re-arrangement. After this initial jump, the rate of change decreases for the rest of the perturbation. Then for the $\sim 10^4$ generations after the perturbation, the rate of divergence rapidly increases. This rate of increase is enough to catch up and overtake the unperturbed systems so that, by the end of the experiment the perturbed systems manage to explore more of the `landscape' \citep{arthur2017decision} and thus reach better final states.

Figure \ref{fig:quakehist}, also shows a histogram of the population difference $\Delta N = N(5\times10^4)-N(10^4)$ for the surviving runs of the $\mu = 0.4$ perturbation \emph{when there is a quake during the perturbation period in the perturbed run and not otherwise}. Operationally, a quake is defined as any change in the core composition together with at least a 5\% change in the total population $N$. The key point about this plot is that is is skewed right, towards higher population excess. This means, when there is a quake or core rearrangement during the perturbation the final population ends up higher than if there was no quake. The other possibilities: quake during the unperturbed and not the perturbed run, both quake and no quake, give symmetric distributions. This means it is runs which have perturbation induced quakes which are responsible for the increase in average population.

Close inspection of the runs which quake during the perturbation, and are responsible for the positive value of $\Delta N$, indicate that the main cause of these is the variance in core species populations. While the population ratios $N_i/N$ are fixed, the absolute populations are much smaller in the perturbation. In equilibrium we have $p(f_i) \simeq p_k$ and the expected number of reproductions of species $i$ in one generation ($N/p_k$ trials) is just the binomial expectation $N_i$. The binomial variance is $N_i (1-p_k)$, and the square root of this measures the average fluctuation size. The signal to noise ratio is then $\sim \sqrt{N_i}$ i.e. when the population is low the variance around the mean value is, relatively, much higher. 

This higher variance in $N_i$ can cause spontaneous core collapse when a fluctuation takes one of the core species to $N_i = 0$. However the most significant effect observed is the translation of fluctuations in $N_i$ to fluctuations in $r$ which can, transiently but significantly, reduce the quake barrier. This makes a quake much more likely to happen, quakes which would not have been possible without the perturbation. During a quake the populations can get very low for a brief period (see Figure \ref{fig:singleruns} for example), making the runs much more likely to go extinct. If a run survives the perturbation but at a worse (lower $N$) qESS it is also much more likely to go extinct. Quakes which cause increases in $N$ are much more likely to survive, and therefore have higher $r$ and so be more stable - this explains the initial jump and then plateau in $d$ in Figure \ref{fig:quakehist}. Runs which don't quake during the perturbation `catch up' with the unperturbed runs after the perturbation goes away and the rate of reproduction increases, which explains the increase in $d$ after the perturbation ends.

\textbf{In summary:} Hostile conditions increase the importance of population fluctuations. This enables more and different quakes - allowing more exploration and enhancing the entropic ratchet mechanism. However quakes are risky, and only those which have positive outcomes survive to be counted - this is selection by survival. This could well have important implications for our search for an inhabited exoplanet and merits further study.

\section{Other Perturbation Experiments}\label{sec:other}

As discussed in the Introduction, `perturbations' of various sizes and severity have affected life on Earth. We don't seek to model any exactly, our model is too conceptual to simulate Earth history, however we can explore some illustrative examples to give us an idea of how the above results are affected by the characteristics of the perturbation and life's response to it.

\subsection{Short Perturbations}

\begin{figure*}
    \centering
    \includegraphics[width=\textwidth]{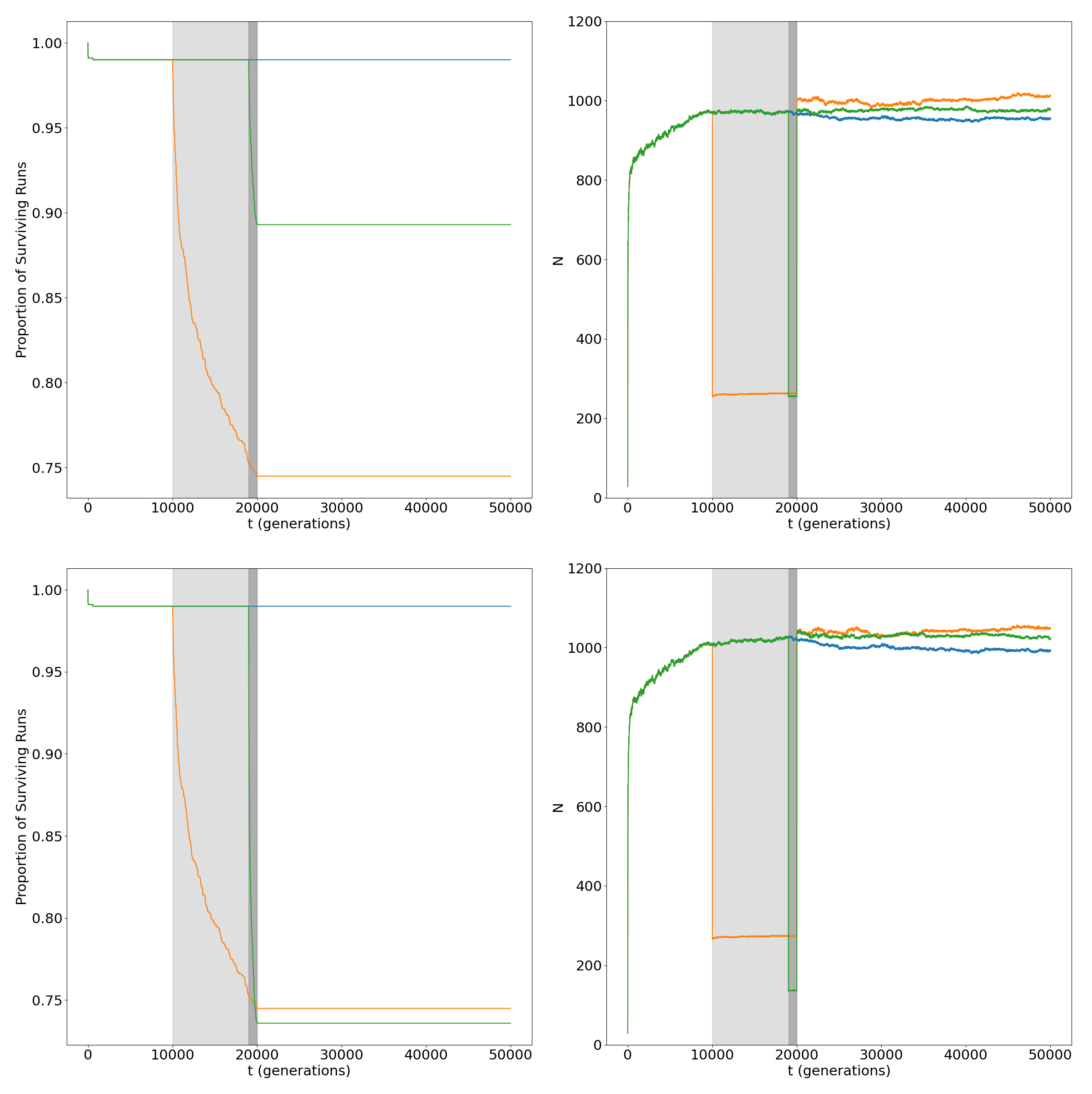}
    \caption{Showing long and short duration perturbations. Only runs which survive all 3 conditions are included in the average. Blue: No perturbation, Orange: long perturbation, Green: short perturbation. Left: Proportion of survivors, Right: Average population. Top: $\mu_{long} \rightarrow 0.4$, $\mu_{short} \rightarrow 0.4$, Bottom: $\mu_{long} \rightarrow 0.4$, $\mu_{short} \rightarrow 0.8$.}
    \label{fig:short}
\end{figure*}

Figure \ref{fig:short} shows the average of 1000 simulations performed as in the previous section along with some new simulations where the perturbations begins later, at $t=19000$, and only lasts for $1000$ generations. When the short perturbation is of the same severity, $\mu$, as the long one, the effect on survival and on $\Delta N$ is less. However when doubling $\mu$ so that the rate of extinction is similar, the effect on $\Delta N$ increases and the two cases are roughly equal. One could try to study the exact dependence of $\Delta N$ on $\mu$ however this depends on $\mu_0$ and a number of other model parameters. The results above are illustrative of the fact that it is the total `intensity' of the perturbation, duration $\times$ severity, that is key.

\subsection{Multiple Perturbations}

\begin{figure*}
    \centering
    \includegraphics[width=\textwidth]{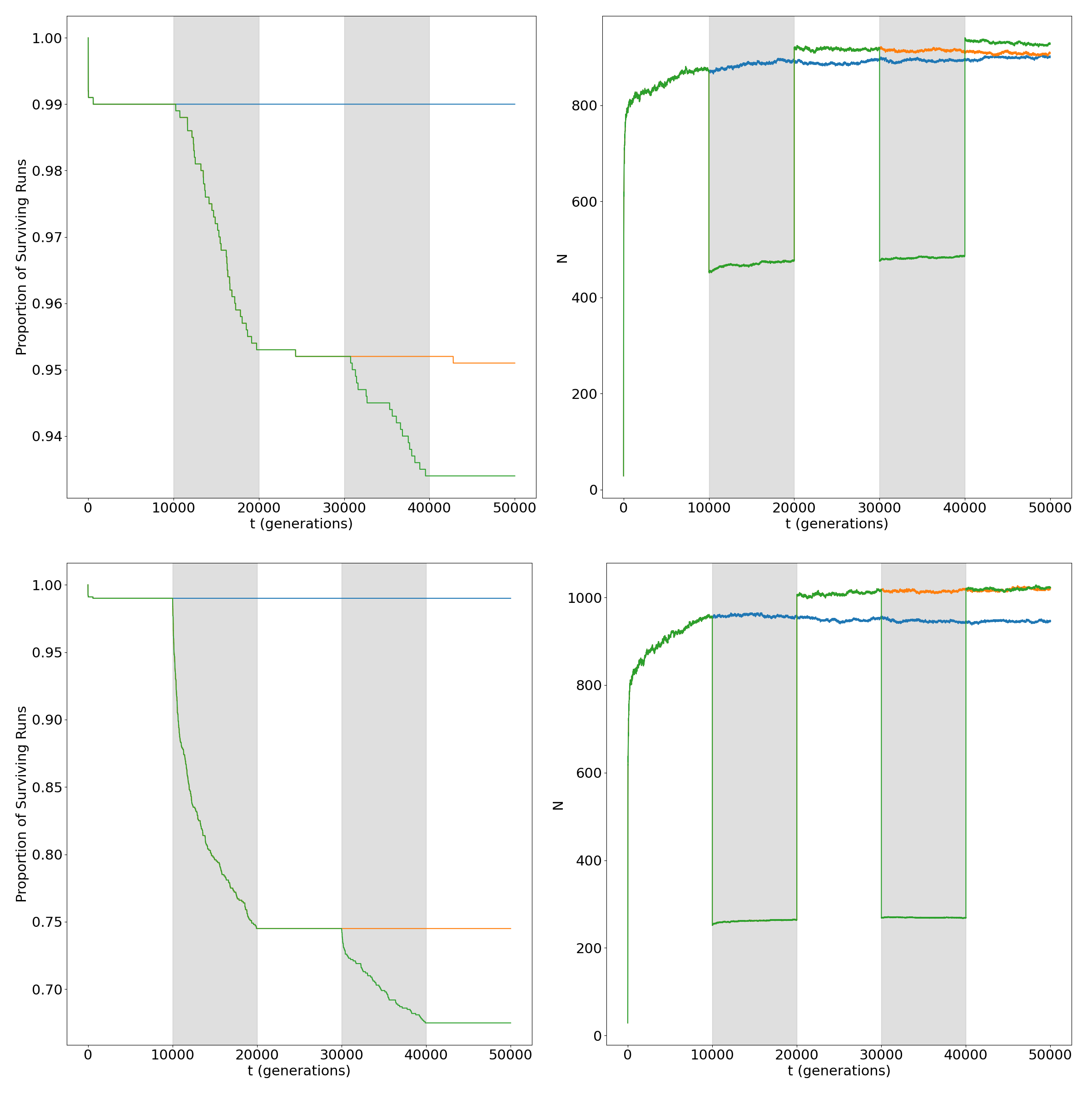}
    \caption{Showing 0, 1 and 2 perturbations. Only runs which survive all 3 conditions are included in the average. Blue: No perturbation, Orange: 1 perturbation, Green: 2 perturbations. Left: Proportion of survivors, Right: Average population. Top: $\mu \rightarrow 0.2$, Bottom: $\mu \rightarrow 0.4$.}
    \label{fig:twice}
\end{figure*}

Figure \ref{fig:twice} shows the effect of a second period of perturbation after the first. We see that for weak perturbations we can get a compounding effect on $\Delta N$. However, for stronger perturbations the second period doesn't have any effect on $\Delta N$. In the former case the second period of perturbation seems to simply increase in perturbation intensity, by effectively increasing the duration. In the latter case the initial perturbation seems to be sufficiently intense (that is severe enough and long enough) to have caused either a jump to a better and  more stable qESS or a total extinction. The second perturbation is then acting on systems which have already been selected at this level and so has little effect, beyond some additional SBS ending more of the runs.

\subsection{Multiple Refugia}

\begin{figure*}
    \centering
    \includegraphics[width=\textwidth]{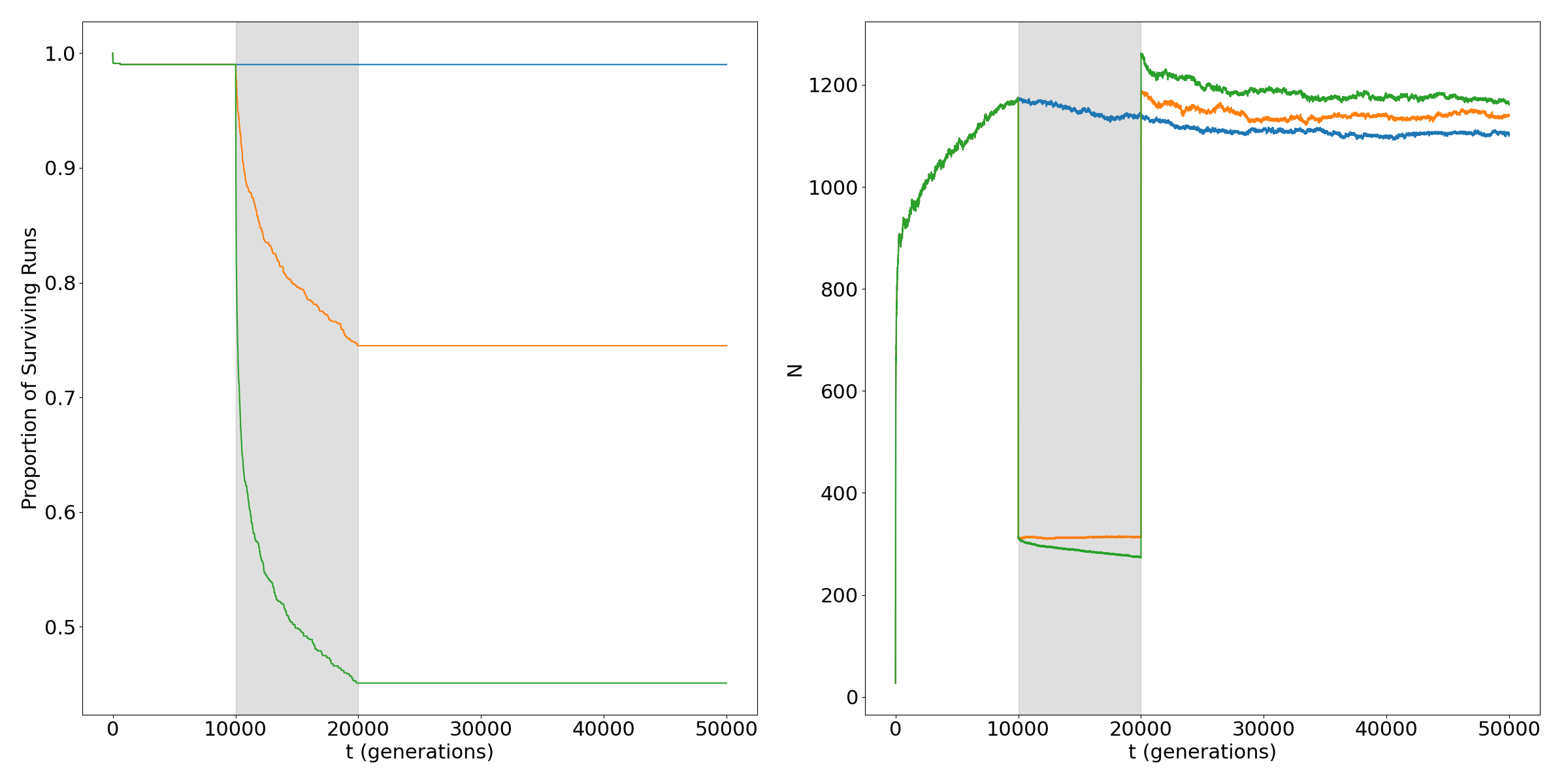}
    \caption{$K=1$ versus $K=4$ refugia. Blue: No perturbation, Orange: 1 refugia $\mu=0.4$, Green: 4 refugia $\mu=1.6$. Left: Proportion of survivors, Right: Average population.}
    \label{fig:manyrefugia}
\end{figure*}

As discussed in Section \ref{sec:refugia}, there are numerous ways in which species attempt to survive periods of stress. We have been studying the case where there is only a single refugia which is hospitable to life during the perturbation. It is also possible that multiple refugia exist and it is interesting to ask what the difference would be between say, an open equatorial ocean (single refugia) or a number of open seas or habitable `patches' of ice (multiple refugia) during a snowball Earth. 

To simulate this we run the TNM for $10^4$ generations, then randomly distribute the individual agents in the system into $K$ refugia. We then run these $K$ systems independently for $10^4$ generations with the perturbation $\mu$. After this we merge all surviving individuals from each of the $K$ refugia back into a single system and run for another $3 \times 10^4$ generations at $\mu_0$. It is obvious that higher $K$ at the same value of $\mu$ gives more chance for runs to survive, quake and so on. Thus using $N \sim \frac{r}{\mu}$ we can `fairly' compare the single refugia case to the $K$ refugia case at values of $\mu$ and $K \mu$ respectively. We are asking if it is better (in terms of survival probability and $\Delta N$) for an ecosystem to have all of its individuals contained within one big refugia or divided into many small ones, during external perturbations.

Figure \ref{fig:manyrefugia} shows one refugia at $\mu=0.4$ compared to four at $\mu=1.6$. We see that both the rate of extinction and the `benefit' $\Delta N$ are increased in the $K=4$ case. $\mu = 1.6$ is quite an extreme perturbation and so many of the runs go extinct, even with $K=4$ chances to survive the perturbation (i.e. four separate refugia). This extreme value of $\mu$ results in lower populations in each of the refugia, and thereby even greater chances of a quake occurring in any of them. Therefore, for runs where at least one of the refugia is inhabited at the end of the perturbation, the increased rate of quakes yield a higher population in the subsequent qESS state. 

\section{Discussion}\label{sec:discussion}

The results presented here should be interpreted carefully. It is true that \emph{surviving} runs which endured the perturbation tend to have higher populations, and all the other Gaian features that this entails within the context of the TNM: stability, diversity, positive species environment coupling etc. see \cite{arthur2022selection}. From the perspective of the surviving population the perturbations are ultimately helpful. However, many result in complete extinction. The idea that large events which are detrimental to carrying capacity might be harmful to life is expected, but the idea that they could be beneficial is not. Large perturbations present both an opportunity and a risk. By weakening the core, new possibilities are opened, at the cost of a significant risk of total extinction. 

In terms of the selection mechanisms discussed in the introduction - both Selection by Survival (SBS) and Entropic Ratcheting (ER) mechanisms are enhanced. SBS simply says any runs which survived the perturbation had to have properties which enabled their survival. In this case they are runs which have higher than average abundance. More subtle, but potentially more interesting, is the fact that runs which survive tend to be better, \emph{because of} the perturbation enhancement to the Entropic Ratchet. 

These effects are really apparent when averaging over many possible life histories. Earth history is only a single time series. There are suggestions, as noted in the Introduction, that large changes in the Earth System are often observed soon after a large perturbation. There is also the opposite observation, the so-called boring billion during the Proterozoic \citep{lenton2013revolutions} was a period of relative stability and slow evolutionary innovation. Thus these ideas have some support in Earth history, and reproducing these mechanisms in such a simplified framework allows us to  understand the potential behaviour of the Earth over its deep past and futute. 

However, where this framework could prove vital is in the application for our search for habitable or indeed inhabited worlds beyond the solar system. Many studies of exoplanets have been performed focused on identifying potentially `habitable' planets through the application of `abiotic' climate models, i.e. neglecting any potential life's impact on the climate. These have also, largely focused on the modern Earth system \citep[e.g.][]{fauchez2022} although work has begun recognising that perhaps a greater probable state would be that of the Archean Earth with its more simple biosphere \citep{arney2016}. However, as discussed in this work, Life on Earth has had a huge impact on the climate \citep[e.g.][]{lenton2013revolutions} and it might be possible that habitable conditions can only persist for long timescales on inhabited planets \citep{goldblatt2016}. Of course, modelling the complex interactions of a distant planetary climate system, including biogeochemical feedbacks from potential life forms is a significant challenge. However, as we detect more and more planets which are designated as potentially habitable we must begin to confront this problem and guide what will be resource intensive follow-up observations to regions of exoplanetary parameter space that we deem most likely to host life. In this regard, simple model frameworks, as independent as possible of the nature of the system itself, are a powerful tool in beginning to map out this likelihood space. With the many thousands of potentially habitable exoplanets likely to exist in our local region of the galaxy alone, it is vital that we attempt to develop a statistical understanding of where we are \emph{most likely} to find life. Our work has shown that the details of the lifeforms relying on a given metabolism are largely unimportant \citep{nicholson2022} and that, perhaps, the edges of the traditional habitable zone \citep{kasting1993} may be more fruitful places to search for long-lived, established life \citep{nicholson2023}. In this work, we further demonstrate that perturbations during the evolution of this life may actually lead to an enhancement in the abundance of life. We have a long way to go and lots of work to complete before we can confidently interpret a potential biosignature from a distant planet, but efforts such as this are vital in beginning to understand \emph{where to look, and what to look for}.

\section*{Data statement}
Code used to generate data is available upon reasonable request from the authors

\section*{Acknowledgements}
This work was partly funded by the Leverhulme Trust through a research project grant [RPG-2020-82] and a UKRI Future Leaders Fellowship [MR/T040866/1].

\bibliographystyle{mnras}
\bibliography{main}

\end{document}